\documentclass[aps,prl,showpacs,twocolumn,floats,epsfig]{revtex4}
\usepackage{amssymb}
\usepackage{amsbsy}
\usepackage{amsmath}
\usepackage{epsfig}

\begin{document}

\title{Tuning Kondo physics in Graphene with gate voltage}

\author{K. Sengupta$^{(1)}$ and G. Baskaran$^{(2)}$}

\affiliation{$^{(1)}$TCMP division, Saha Institute of Nuclear
Physics, 1/AF Bidhannagar, Kolkata-700064, India. \\ $^{(2)}$ The
Institute of Mathematical Sciences, C.I.T Campus, Chennai-600113,
India.}

\date{\today}

\begin{abstract}

We show theoretically that graphene, which exhibits a massless Dirac
like spectrum for its electrons, can exhibit unconventional Kondo
effect that can be tuned by an experimentally controllable applied
gate voltage. We demonstrate the presence of a finite critical Kondo
coupling strength in neutral graphene. We discuss the possibility of
multichannel Kondo effect in this system which might lead to a
non-Fermi liquid like ground state and provide a discussion of
possible experimental realization of Kondo phenomenon in graphene.

\end{abstract}

\maketitle

Graphene, a two-dimensional single layer of graphite, has been
recently fabricated by Novoselov {\it et.\,al.} \cite{nov1}. In
graphene, the energy bands touch the Fermi energy at six discrete
points at the edges of the hexagonal Brillouin zone. Out of these
six Fermi points, only two are inequivalent; they are commonly
referred to as $K$ and $K'$ points \cite{ando1}. The quasiparticle
excitations about these $K$ and $K'$ points obey linear Dirac-like
energy dispersion. The presence of such Dirac-like quasiparticles is
expected to lead to a number of unusual electronic properties in
graphene including relativistic quantum hall effect with unusual
structure of Hall plateaus \cite{shar1}. Recently, experimental
observation of the unusual plateau structure of the Hall
conductivity has confirmed this theoretical prediction \cite{nov2}.
Further, the presence of Dirac-like quasiparticles in graphene
provides us with an experimental test bed for Klein paradox
\cite{nov2}, transmission resonance \cite{sengupta1} and
Lorenz-boost \cite{vinu} type phenomena.

An extremely interesting phenomenon in conventional metal systems is
the Kondo effect which occurs in the presence of dilute
concentration of localized quantum spins coupled to the
spin-degenerate Fermi sea of metal electrons \cite{kondoref1}. The
impurity spin-electron interaction then results in perfect or
partial screening of the impurity spin leading to an apparently
divergent resistance, as one approaches zero temperature. It also
results in a sharp `Kondo Resonance' in electron spectral functions.
Recent developments in quantum dots and nano devices have given new
ways in which various theoretical results in Kondo physics, which
are not easily testable otherwise, can be tested and confirmed
experimentally \cite{kondoref2}. Most of the early studies in Kondo
effect were carried on for conventional metallic systems with
constant density of states (DOS) at the Fermi surface
\cite{affleck1}. Some studies on Kondo effect in possible flux
phases \cite{frad1}, nodal quasiparticles in d-wave superconductors
\cite{subir1}, Luttinger liquids \cite{furu1}, and hexagonal Kondo
lattice \cite{saremi}, for which the DOS of the associated Fermions
vanishes as some power law at the Fermi surface, has also been
undertaken. However, although effect of non-magnetic impurities has
been studied \cite{bena1}, there has been no theoretical study till
date on the nature of Kondo effect in graphene.

In this letter we present a large $N$ analysis for a generic local
moment coupled to Dirac electrons in graphene to show that Kondo
effect in graphene is unconventional can be tuned by gate voltage.
We demonstrate the presence of a finite critical Kondo coupling
strength in neutral graphene. We point out that local moments in
graphene can lead to non Fermi-liquid ground state via multi channel
Kondo effect. We also suggest possible experimental realization of
such Kondo scatterers in graphene.

The crucial requirement for occurrence of Kondo effect is that the
embedded impurities should retain their magnetic moment in the
presence of conduction of electrons of graphene. We will not
quantitatively address the problem of local moment formation in the
presence of Dirac sea of electrons in graphene in the present paper.
We expect that large band width and small linearly vanishing density
of states at the fermi level in graphene should make survival of
impurity magnetic moment easier than in the conventional 3D metallic
matrix. A qualitative estimate of the resultant Kondo coupling can
be easily made considering hybridization of electrons in $\pi$ band
in graphene with $d$ orbitals of transition metals. Typical hopping
matrix elements for electrons in $\pi$ band is $t \sim 2$eV and
effective Hubbard $U$ in transition metals is $8$eV. So the Kondo
exchange $J \sim 4t^2/U$, estimated via standard Schrieffer-Wolf
transformation, can be as large as $2$ eV which is close to one of
the largest $J \simeq 2.5$ eV for Mn in Zn. In the rest of this
work, we shall therefore use the Kondo Hamiltonian \cite{kondoref3}
as our staring point.

Our analysis begins with the Hamiltonian for non-interacting Dirac
electron in graphene. In the presence of a gate voltage $V$, the
Hamiltonian can be expressed in terms of electron annihilation
operators $\Psi_{A(B) \alpha}^s$ at sublattice $A(B)$ and Dirac
point $s=K,K$ with spin $\alpha=\uparrow,\downarrow$ as
\begin{eqnarray}
H &=& \int \frac{d^2 k}{(2\pi)^2} \left( \Psi_{A \alpha}^{s \dagger}({\bf k}),
\Psi_{B \alpha}^{s\dagger}({\bf k})\right) \nonumber\\
&& \times \left(\begin{array}{cc} {eV \quad \hbar v_F (k_x-i {\rm sgn}(s) k_y)} \\
{\hbar v_F(k_x+i {\rm sgn}(s) k_y) \quad  eV} \end{array}\right)
\left(
\begin{array}{c} {\Psi_{A \alpha}^{s}({\bf
k})}\\{\Psi_{B\alpha}^{s}({\bf k})}
\end{array}\right) \label{ke1}
\end{eqnarray}
where ${\rm sgn}(s)=1(-1)$ for $s=K(K')$, $v_F$ is the Fermi
velocity of graphene, and all repeated indices are summed over. In
Eq.\ \ref{ke1} and in rest of the work, we shall use an upper
momentum cutoff $k_c = \Lambda/(\hbar v_F)$, where $\Lambda \simeq
2$eV corresponds to energy up to which the linear Dirac dispersion
is valid, for all momenta integrals.

Eq.\ \ref{ke1} can be easily diagonalized to obtain the eigenvalues
and eigenfunctions of the Dirac electrons: $E_{\pm} = eV \pm \hbar
v_F k$ where ${\bf k} = (k_x,k_y)=(k,\theta)$ denote momenta in
graphene and $(u_A^{s \pm},u_B^{s \pm}) = 1/{\sqrt{2}} \left(1, \pm
\exp \left(i {\rm sgn}(s) \theta\right) \right)$. Following Ref.\
\cite{frad1}, we now introduce the $\xi$ fields, which represents
low energy excitations with energies $E_{\pm}$, and write
\begin{eqnarray}
\Psi_{A\alpha}^{s}({\bf k}) &=& \sum_{j=\pm} u^{sj}_A
\xi_{j\alpha}^s = 1/\sqrt{2} (\xi_{+ \alpha}^s ({\bf k})+ \xi_{-
\alpha}^s ({\bf k})), \nonumber\\
\Psi_{B\alpha}^{s}({\bf k}) &=& \exp(i\theta)/\sqrt{2} (\xi_{+
\alpha}^s ({\bf k}) - \xi_{- \alpha}^s ({\bf k})).   \label{xifef}
\end{eqnarray}

In what follows, we shall consider a single impurity to be centered
around ${\bf x}=0$. Thus to obtain an expression for the coupling
term between the local moment and the conduction electrons, we shall
need to obtain an expression for $\Psi({\bf x}=0)\equiv \Psi(0)$. To
this end, we expand the $\xi$ fields in angular momentum channels
$\xi_{+ \alpha}^s ({\bf k})=\sum_{m=-\infty}^{\infty} e^{i m \theta}
\xi_{+ \alpha}^{m s} (k)$, where we have written ${\bf
k}=(k,\theta)$. After some straightforward algebra, one obtains
\begin{eqnarray}
\Psi_{B\alpha}^{s}(0)&=& \frac{1}{\sqrt{2}} \int_0^{k_c} \frac{k
dk}{2\pi} \left( \xi_{+ \alpha}^{-{\rm sgn}(s) s} (k) -
\xi_{- \alpha}^{-{\rm sgn}(s) s} (k)\right), \nonumber\\
\Psi_{A\alpha}^{s}(0)&=& \frac{1}{\sqrt{2}} \int_0^{k_c} \frac{k
dk}{2\pi} \left( \xi_{+ \alpha}^{0 s} (k) + \xi_{- \alpha}^{0 s}
(k)\right). \label{psif}
\end{eqnarray}
Note that $\Psi_B(0) $ receives contribution from $m=\pm 1$ channel
while for $\Psi_A(0)$, the $m=0$ channel contributes. The Kondo
coupling of the electrons with the impurity spin is given by
\begin{eqnarray} H_K &=& \frac{g}{2k_c^2} \sum_{s=1}^{N_s}
\sum_{l=1}^{N_f} \sum_{\alpha,\beta =1}^{N_c}  \sum_{a=1}^{N_c^2-1}
\Psi^{s \,\dagger}_{l \alpha}(0) \tau_{\alpha \beta}^a  \Psi^{s}_{l
\beta}(0) S^a, \label{coupling1}
\end{eqnarray}
where $g$ is the effective Kondo coupling for energy scales up to
the cutoff $\Lambda$, ${\bf S}$ denotes the spin at the impurity
site, ${\bf \tau}$ are the generators of the SU($N_c$) spin group,
and we have now generalized the fermions, in the spirit of large $N$
analysis, to have $N_s$ flavors (valley indices) $N_f$ colors
(sublattice indices) and $N_c$ spin. For realistic systems
$N_f=N_c=N_s=2$. Here we have chosen Kondo coupling $g$ to be
independent of sublattice and valley indices. This is not a
necessary assumption. However, we shall avoid extension of our
analysis to flavor and/or color dependent coupling term for
simplicity. Also, the Dirac nature of the graphene conduction
electrons necessitates the Kondo Hamiltonian to mix $m=\pm 1$ and
$m=0$ channels (Eqs.\ \ref{psif} and \ref{coupling1}). This is in
complete contrast to the conventional Kondo systems where the Kondo
coupling involves only $m=0$ angular momentum channel.

The kinetic energy of the Dirac electrons can also be expressed in
terms of the $\xi$ fields: $H_0 = \int_0^{\infty} \frac{k dk}{2\pi}
\sum_{m=-\infty}^{\infty} \sum_{s,\alpha} \left(E_{+}(k)
\xi_{+\alpha}^{m s \, \dagger} \xi_{+ \alpha}^{m s} + E_{-}(k)
\xi_{-\alpha}^{m s\, \dagger} \xi_{- \alpha}^{m s} \right)$.
Typically such a term involves all angular momenta channels. For our
purpose here, it will be enough to consider the contribution from
electrons in the $m=0,\pm 1$ channels which contribute to scattering
from the impurity (Eqs.\ \ref{psif} and \ref{coupling1}). To make
further analytical progress, we now unfold the range of momenta $k$
from $(0, \infty)$ to $(-\infty,\infty)$ by defining the fields
$c_{1(2) \alpha}^{s}$
\begin{eqnarray}
c_{1(2) \alpha}^{s}(k) &=& \sqrt{\left|k\right|}
\xi_{+ \alpha}^{0(-{\rm sgn}(s)) s} (|k|), \quad k>0 ,\nonumber\\
c_{1(2) \alpha}^{s}(k) &=& +(-) \sqrt{\left|k\right|} \xi_{-
\alpha}^{0(-{\rm sgn}(s)) s} (|k|), \quad k < 0, \label{cfield}
\end{eqnarray}
so that one can express the $\Psi$ fields as $\Psi_{A(B)
\alpha}^s(0) = \int_{-\infty}^{\infty} \frac{ dk}{2\pi}
\sqrt{\left|k\right|} c_{1(2) \alpha}^s (k)$. In terms of the
$c_{1(2) \alpha}^{s}$ fields, the kinetic energy (in the $m=0,\pm 1$
channels) and the Kondo terms in the Hamiltonian can therefore be
written as
\begin{eqnarray}
H_0 &=&  \int_{-k_c}^{k_c} dk/(2\pi)  E_k
c_{l \alpha}^{s\, \dagger} c_{l \alpha}^{s} \nonumber\\
H_K &=& g/(8 \pi^2 k_c^2) \int_{-k_c}^{k_c} \int_{-k_c}^{k_c}
\sqrt{\left|k\right|} \sqrt{\left|k'\right|}
dk dk' \nonumber\\
&& \times \left( c_{l \alpha}^{s\, \dagger} (k) \, \tau_{\alpha
\beta}^a \, c_{l \beta }^{s}(k') \, S^a \right), \label{coupling2}
\end{eqnarray}
where $E_k = eV + \hbar v_F k$ and summation over all repeated
indices are assumed.

Next we follow standard procedure \cite{newns1} of representing the
local spin by SU($N_c$)Fermionic fields $f_{\alpha}$ and write the
partition function of the system in terms of the $f$ and $c$ fields
\begin{eqnarray}
Z &=& \int {\mathcal D} c {\mathcal D} c^{\dagger}  {\mathcal D} f
{\mathcal D} f^{\dagger}
{\mathcal D} \epsilon  \, e^{-S/\hbar}, \quad S = S_0 + S_1 + S_2 \nonumber \\
S_0 &=& \int_0^{\beta\hbar } d \tau \int_{-k_c}^{k_c} dk/(2 \pi)
\left( c_{l \alpha}^{s\, \dagger}(k,\tau)
G_0^{-1}  c_{l \alpha}^{s} (k,\tau)\right), \nonumber\\
S_1 &=& J/(4 \pi^2 N_c k_c^2) \int_0^{\beta \hbar} d \tau
\int_{-k_c}^{k_c} \int_{-k_c}^{k_c} \sqrt{\left|k\right|}
\sqrt{\left|k'\right|} dk dk' \nonumber\\
&& \times \left[c_{l \alpha}^{s\, \dagger} (k,\tau) \, \tau_{\alpha
\beta}^a \, c_{l \beta }^{s}(k',\tau) f_{\gamma}^{\dagger}(\tau)
\tau_{\gamma \delta}^a f_{\delta}(\tau) \right] \nonumber \\
S_2 &=& \int_0^{\beta \hbar} d\tau  \left[\left(
f_{\alpha}^{\dagger}(\tau)\left[\hbar \partial_{\tau}+\epsilon(\tau)
\right] f_{\alpha} (\tau) \right)- \epsilon(\tau) Q \right],
\label{s2}
\end{eqnarray}
where $G_0^{-1} = \hbar \partial_{\tau} +E_k$ is the propagator for
$c$ fields, $J= g N_c/2$ is the renormalized Kondo coupling, we have
imposed the impurity site occupancy constraint $\sum_{\alpha}
f_{\alpha}^{\dagger} f_{\alpha} = Q $ using a Lagrange multiplier
field $\epsilon(\tau)$.

We now use the identity $\tau_{\alpha \beta}^a \tau_{\gamma
\delta}^a = N_c \delta_{\alpha \delta} \delta_{\beta \gamma} -
\delta_{\alpha \beta} \delta_{\gamma \delta}$ \cite{newns1} and
decouple $S_1$ using a Hubbard-Stratonovitch field $\phi_l^s$. In
the large $N_c$ limit one has $S= S_0 + S_2 + S_3 + S_4$, where
\begin{eqnarray}
S_3 &=& \int_0^{\beta \hbar} d \tau \int_{-k_c}^{k_c}
\frac{\sqrt{\left|k\right|} dk }{(2\pi)} \left(
\phi_l^{\ast\,s}(\tau) c_{l \alpha}^{s\, \dagger} (k,\tau)
f_{\alpha}(\tau) + {\rm h.c} \right)
\nonumber \\
S_4 &=& N_c k_c^2/J  \int_0^{\beta \hbar} d \tau
\phi_l^{\ast\,s}(\tau)  \phi_l^s(\tau) . \label{s4}
\end{eqnarray}
Note that at the saddle point level $ \left<\phi_l^s\right>  \sim
\left< \sum_{\alpha} c_{l\alpha} ^{s\,\dagger} f_{\alpha}\right>$ so
that a non-zero value of $\phi_l^s$ indicates the Kondo phase. In
what follows, we are going to look for the static saddle point
solution with $\phi_l^s (\tau) \equiv \phi_0$ and $\epsilon(\tau)
\equiv \epsilon_0$ \cite{newns1}. In this case, it is easy to
integrate out the $c$ and $f$ fields, and obtain an effective action
in terms of $\phi_ 0$ and $\epsilon_0$ and one gets $S' = S_5 + S_6$
with
\begin{eqnarray}
S_5 &=& -\beta \hbar N_c {\rm Tr} \left[\ln \left(i \hbar \omega_n
-\epsilon_0 - N_s N_f \phi_0^{\ast} G'_0(i\omega,V) \phi_0 \right)
\right],
\nonumber \\
S_6 &=& \beta \hbar \left( N_s N_c N_f k_c^2 \left|\phi_0\right|^2/J
- \epsilon_0 Q \right), \label{s6}
\end{eqnarray}
where ${\rm Tr}$ denotes Matsubara frequency sum as well as trace
over all matrices and the Fermion Green function $G'_0(ip_n,q)
\equiv G'_0$ is given by \cite{frad1}
\begin{eqnarray}
G'_0 &=& \frac{-\Lambda}{2 \pi (\hbar v_F)^2} (ip_n-q) \ln
\left[1/\left|ip_n-q\right|^2\right], \label{g0}
\end{eqnarray}
where, in the last line we have switched to dimensionless variables
$p_n=\hbar \omega_n/\Lambda$ and $q = eV/\Lambda$.

One can now obtain the saddle point equations from Eq.\ \ref{s6}
which are given by $\delta S'/\delta \phi_0 =0$ and $\delta
S'/\delta \epsilon_0 =0$. Using Eqs.\ \ref{s6} and \ref{g0}, one
gets (after continuing to real frequencies and for $T=0$)
\begin{eqnarray}
1/J &=& - \Lambda/(\pi \hbar v_F k_c^2 )^2 \int_{-1}^{0} dp\,
G_0(p - \nu - \Delta_0 G_0 /2)^{-1}, \nonumber\\
Q/N_c &=& 1/(2 \pi) \int_{-1}^{0} dp\, \nu (p - \nu - \Delta_0 G_0
/2)^{-1}, \label{saddle}
\end{eqnarray}
where we have defined the dimensionless variable $\Delta_0 = N_f N_s
|\phi_0|^2/(\pi \hbar^2 v_F^2)$, $p = \hbar \omega/\Lambda$, $G_0 =
2 \pi (\hbar v_F)^2 G'_0/\Lambda$, $\nu= \epsilon_0/\Lambda \ge 0$,
and have used the energy cutoff $\Lambda$ for all frequency
integrals. At the critical value of the coupling strength, putting
$\nu=0$ and $\Delta_0=0$, we finally obtain the expression for
$J_c(q,T)$
\begin{eqnarray}
J_c(q,T) &=& J_c(0) \left[1-2q\ln\left(1/q^2\right) \ln\left(k_B
T/\Lambda\right) \right]^{-1} \label{criticalJ}
\end{eqnarray}
where the temperature $k_B T$ is the infrared cutoff, $J_c(0) = (\pi
\hbar v_F k_c^2)^2/\Lambda = \pi^2 \Lambda$ is the critical coupling
in the absence of the gate voltage, and we have omitted all
subleading non-divergent term which are not important for our
purpose. For $V=0=q$, we thus have, analogous to the Kondo effect in
flux phase systems \cite{frad1}, a finite critical Kondo coupling
$J_c(0) = \pi^2 \Lambda \simeq 20$eV which is a consequence of
vanishing density of states at the Fermi energy for Dirac electrons
in graphene. Of course, the mean-field theory overestimates $J_c$. A
quantitatively accurate estimate of $J_c$ requires a more
sophisticated analysis which we have not attempted here.

\begin{figure}
\rotatebox{0}{
\includegraphics*[width=\linewidth]{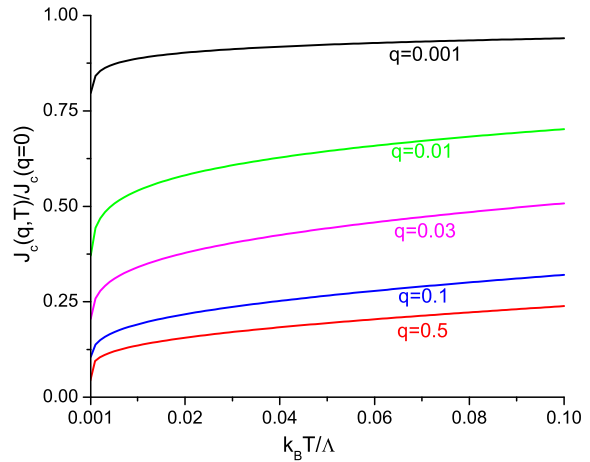}}
\caption{Sketch of the critical Kondo coupling $J_c(q,T)$ as a
function of temperature for several applied voltages $q=eV/\Lambda$.
The Kondo phase exists for $J > J_c$.} \label{fig1}
\end{figure}
The presence of a gate voltage leads to a Fermi surface and
consequently $J_c(q,T) \rightarrow 0$ as $T \rightarrow 0$. For a
given experimental coupling $J < J_c(0)$ and temperature $T$, one
can tune the gate voltage to enter a Kondo phase. Fig.\ \ref{fig1},
which shows a plot of $J_c(q,T)$ as a function of $T$ for several
gate voltages $q$ illustrates this point. The temperature
$T^{\ast}(q)$ below which the system enters the Kondo phase for a
physical coupling $J$ can be obtained using $J_c(q,T^{\ast})=J$
which yields
\begin{eqnarray}
k_B T^{\ast} &=& \Lambda
\exp\left[(1-J_c(0)/J)/(2q\ln[1/q^2])\right]
\end{eqnarray}
For a typical $J \simeq 2$eV and voltage $eV \simeq 0.5$eV,
$T^{\ast} \simeq 35$K \cite{comment1}. We stress that even with
overestimated $J_c$, physically reasonable $J$ leads to
experimentally achievable $T^{\ast}$ for a wide range of
experimentally tunable gate voltages.

We now discuss the possible ground state in the Kondo phase
qualitatively. In the absence of the gate voltage a finite $J_c$
implies that the ground state will be non-Fermi liquid as also noted
in Ref.\ \cite{frad1} for flux phase systems. In view of the large
$J_c$ estimated above, it might be hard to realize such a state in
undoped graphene. However, in the presence of the gate voltage, if
the impurity atom generates a spin half moment and the Kondo
coupling is independent of the valley(flavor) index, we shall have a
realization of two-channel Kondo effect in graphene owing to the
valley degeneracy of the Dirac electrons. This would again lead to
overscreening and thus a non Fermi-liquid like ground state
\cite{affleck1}. The study of details of such a ground state
necessitates an analysis beyond our large $N$ mean-field theory. To
our knowledge, such an analysis has not been undertaken for Kondo
systems with angular momentum mixing. In this work, we shall be
content with pointing out the possibility of such a multichannel
Kondo effect in graphene and leave a more detailed analysis as an
open problem for future work.

Next, we discuss experimental observability of the Kondo phenomena
in graphene. The main problem in this respect is creation of local
moment in graphene. There are several routes to solving this
problem. i) Substitution of a carbon atom by a transition metal
atom. This might in principle frustrate the strong sp$^2$ bonding
and thus locally disturb the integrity of graphene atomic net.
However, nature has found imaginative ways of incorporating
transition metal atoms in p-$\pi$ bonded planar molecular systems
such as porphyrin \cite{por1}. Similar transition metal atom
incorporation in extended graphene, with the help of suitable
bridging atoms, might be possible. ii) One can try chemisorption of
transition metal atoms such as Fe on graphene surface through sp-d
hybridization in a similar way as in intercalated graphite
\cite{chem1}. iii) It might be possible to chemically bond molecules
or free radicals with magnetic moment on graphene surface as
recently done with cobalt pthalocyanene (CoPc) molecule on AU(111)
surface \cite{copc}. This might result in a strong coupling between
graphene and impurity atom leading to high Kondo temperatures as
seen for CoPc on AU(111) surface ($T_K \simeq 280K$). iv) Recently
ferromagnetic cobalt atom clusters with sub nano-meter size,
deposited on carbon nanotube, have exhibited Kondo
resonance\cite{cluster1}. Similar clusters deposition in graphene
might be a good candidate for realization of Kondo systems in
graphene. v) From quantum chemistry arguments, a carbon vacancy, or
substitution of a carbon atom by a boron or nitrogen might lead to a
spin-half local moment formation. In particular, it has been shown
that generation of local defects by proton irradiation can create
local moments in graphite \cite{defect1}. Similar irradiation
technique may also work for graphene.

For spin one local moments and in the presence of sufficiently large
voltage and low temperature, one can have a conventional Kondo
effect in graphene. The Kondo temperature for this can be easily
estimated using $k_BT_K \sim D \exp(-1/\rho J )$ where the band
cutoff $D \simeq10$eV, $J \simeq 2-3$eV  and DOS per site in
graphene $\rho \simeq 1/20$ per eV. This yield $T_K \simeq 6-150$K.
The estimated value of $T_K$ has rather large variation due to
exponential dependence on $J$. However, we note that Kondo effect
due to Cobalt nano-particle in graphitic systems such as carbon
nanotube leads to a high $T_K \approx 50 K$ which means that a large
$J$ may not be uncommon in these systems.

Finally, we note that recent experiments have shown a striking
conductance changes in carbon nanotubes and graphene, to the extent
of being able to detect single paramagnetic spin-half NO$_2$
molecule \cite{novo1}. This has been ascribed to conductance
increase arising from hole doping (one electron transfer from
graphene to NO$_2$). Although Kondo effect can also lead to
conductance changes, in view of the fact that a similar effect has
been also seen for diamagnetic NH$_3$ molecules, the physics in
these experiments is likely to be that of charge transfer and not
local moment formation.

In conclusion, we have shown that the Kondo effect in graphene is
unconventional and can be tuned by an applied gate voltage. We have
shown that it is possible to have multichannel Kondo effect in
graphene and discussed experimental possibilities of its
realization.

After submission of the first version of this manuscript in the
arXiv (arXiv:0705.0257, v1), we became aware of Ref.\ \cite{g1} with
similar conclusion regarding existence of finite critical Kondo
coupling in neutral graphene.


\vspace{-0.8 cm}

\begin{thebibliography}{99}
\bibitem{nov1} K.S. Novoselov {\it et.al.} Science {\bf 306}, 666
(2004).

\bibitem{ando1} T. Ando, J. Phys. Soc. Jpn. {\bf 74} 777 (2005).

\bibitem{shar1} V.P. Gusynin and S.G. Sharapov, Phys. Rev. Lett. {\bf 95},
146801 (2005); N.M.R Peres, F. Guinea, and A. Castro Neto, Phys.
Rev, B {\bf 73}, 125411 (2006).

\bibitem{nov2} K.S. Novoselov {\it et.al.} Nature {\bf 438}, 197
(2006); Y. Zhang {\it et.al.} Nature {\bf 438}, 201 (2005); M.I.
Katsnelson {\it et.al.} Nature Phys. {\bf 2} 177 (2006).

\bibitem{sengupta1}  S. Bhattacharya and K. Sengupta, Phys. Rev.
Lett. {\bf 97}, 217001 (2006); K. Sengupta, cond-mat/0611614
(unpublished).

\bibitem{vinu} V. Lukose, R. Shankar and G. Baskaran,
Phys. Rev. Lett., {\bf 98} 116802 (2007)

\bibitem{kondoref1} See for example A.C. Hewson {\it The Kondo Problem to
Heavy Fermions}, Cambridge University Press (1993).

\bibitem{kondoref2}M. Pustilnik and L. Glazman, J. Phys.:Condens. Matter,
{\bf 16} R 513 (2004);

\bibitem{affleck1} For a review see I. Affleck,
Acta Phys.Polon. {\bf B26} 1869 (1995); cond-mat/9512099;


\bibitem{frad1} C.R. Cassanello and E. Fradkin Phys. Rev. B {\bf
53}, 15079 (1996); D. Withoff and E. Fradkin, Phys. Rev. Lett {\bf
64}, 1835 (1990); K. Insergent, Phys. Rev. B {\bf 54}, 11396 (1996).

\bibitem{subir1} A. Polkovnikov, S. Sachdev and M. Vojta, Phys. Rev.
Lett {\bf  86}, 296 (2001).

\bibitem{furu1} A. Furusaki and N. Nagaosa, Phys. Rev.
Lett {\bf  72}, 892 (1994).

\bibitem{saremi} S. Saremi and P.A. Lee, cond-mat/0610273 (unpublished)

\bibitem{bena1} C. Bena and S. Kivelson, Phys. Rev. B, {\bf 72} 125432
(2005); T.O.  Wehling {\it et al.}, cond-mat/0609503 (unpublished).

\bibitem{kondoref3} J. Kondo, Prog. Theor. Phys. {\bf 32}, 37
(1964).

\bibitem{newns1} N. Read and C.J. Newns, J. Phys. C {\bf 16}, 3273
(1983).

\bibitem{comment1} Note that $J$ in our analysis is an effective
coupling valid below the scale $\Lambda$ ${\it i.e.}$ $J>J_{\rm
bare}$. This might further enhance $T^{\ast}$.


\bibitem{por1} Porphyrin Hand Book, Eds. K. M. Kadish, K.M. Smith and
R. Guilard, (Academic Press, New York) 1999

\bibitem{chem1} M.E. Vol'pin and Yu.N. Novikov, Pure and Appl. Chem.,
vol. {\bf 60}, No. 8, 1133 (1988)

\bibitem{copc} A. Zhao {\it et\,al.}, Science {\bf 309}. 1542
(2005).

\bibitem{cluster1} T.W. Odom et al., Science, {\bf 290} 1459 (2000)


\bibitem{defect1}P.O. Lehtinen {\it et\,al.}, Phys. Rev. Lett {\bf
93}, 187202 (2004).

\bibitem{novo1} T.O. Wehling {\it et\,al.}, cond-mat/0703390
(unpublished); S. Adam, S. Das Sarma and A.K. Geim, cond-mat/0610834
(unpublished).

\bibitem{g1} M. Hentschel and F. Guinea, arXiv:0705.0522.



\end{thebibliography}
\end{document}